\documentclass[a4paper,twocolumn,11pt,unpublished]{quantumarticle}
\pdfoutput=1
\usepackage{braket}
\usepackage{amsthm}
\usepackage{booktabs}
\usepackage{subfig}
\usepackage{graphicx}
\usepackage{epstopdf}
\usepackage{seqsplit}
\usepackage{bm}
\usepackage{amssymb}
\usepackage{stfloats}
\usepackage{tabularx} 

\usepackage[utf8]{inputenc}
\usepackage[english]{babel}
\usepackage[T1]{fontenc}
\usepackage{amsmath}
\usepackage[backref]{hyperref}
\usepackage{float}
\usepackage{tikz}
\usepackage{lipsum}
\makeatletter

\renewcommand{\maketag@@@}[1]{\hbox{\m@th\normalsize\normalfont#1}}%

\makeatother

\newtheorem{remark}{remark}
\begin{document}
	
	\title{Quantum Krylov-Subspace Method Based Linear Solver}
	
	\author{Rui-Bin Xu}
	\affiliation{School of Mathematics, South China University of Technology,  Guangzhou 510640, P.R. China}
	\author{Zhu-Jun Zheng}
	\affiliation{School of Mathematics, South China University of Technology,  Guangzhou 510640, P.R. China}
	\author{Zheng Zheng $^{\dagger}$}
	\affiliation{School of Mathematics, Guangdong University of Education, Guangzhou 510303, P.R. China}
	\email{zhengzhengyv@126.com}
	\maketitle
	\mathchardef\mhyphen="2D
	\begin{abstract}
		Despite the successful enhancement to the Harrow-Hassidim-Lloyd algorithm by Childs et al., who introduced the Fourier approach leveraging linear combinations of unitary operators, our research has identified non-trivial redundancies within this method.  This finding points to a considerable potential for refinement. In this paper, we propose the quantum Krylov-subspace method (QKSM), which is a hybrid classical-quantum algorithm, to mitigate such redundancies. By integrating QKSM as a subroutine, we introduce the quantum Krylov-subspace method based linear solver that not only reduces computational redundancy but also enhances efficiency and accuracy. Extensive numerical experiments, conducted on systems with dimensions up to $2^{10} \times 2^{10}$, have demonstrated a significant reduction in computational resources and have led to more precise approximations.
	\end{abstract}
	\section{Introduction}
	
	\par Over the past decade, the field of quantum computing has advanced rapidly, driven by developments in theoretical understanding and practical applications\cite{zeguendry2023quantum}. Quantum algorithms\cite{arute2019quantum}, based on the properties of quantum mechanics, offer the promise of exponentially faster computation for certain problems compared to classical algorithms. From factoring large integers\cite{shor1999polynomial} to optimizing complex systems\cite{abel2022quantum}, quantum algorithms have the potential to revolutionize various domains of computation.
	\par In the realm of quantum computation, the Quantum Linear Systems Problem (QLSP) is conventionally defined as follows: Given a system of linear equations $\mathcal{H}\bm{x}= \bm{b}$, 
	where $\mathcal{H}\in{\mathbb{C}}^{N\times N}$ and $\bm{b}\neq{\mathbf{0}}\in{\mathbb{C}}^{N}$, the goal is to prepare a quantum state $\ket{\Tilde{\bm{x}}}$ that is proportional to the solution $\bm{x}$ with an error of at most $\epsilon$. Following common practice, $\mathcal{H}$ is typically postulated to be $d$-sparse, Hermitian, with a condition number of $\kappa$, normalized to have a spectral norm of 1 and the eigenvalues of $\mathcal{H}$ fall within the interval $[-1,-1/\kappa]\cup[1/\kappa, 1]$. Moreover, it is widely accepted that the quantum state $\ket{\bm{b}}$ proportional to $\bm{b}$ can be efficiently prepared with a quantum procedure $\mathcal{P}$.
	
	\par In quantum computation research, the QLSP has garnered substantial attention, prompting numerous studies to address this challenge.  The Harrow-Hassidim-Lloyd (HHL) algorithm \cite{PhysRevLett.103.150502} has achieved an $\epsilon$-approximation to the desired quantum state with $\text{poly}(\log N, 1/\epsilon, d, \kappa)$ resources. Among the enhancement of the HHL\cite{PhysRevLett.122.060504,doi:10.1137/16M1087072,bravo2023variational,ambainis2010variable}, Childs et al. \cite{doi:10.1137/16M1087072} introduced the Fourier approach that leverages linear combinations of unitary operators(LCU), demonstrating polylogarithmic scaling in $1/\epsilon$. Despite the Fourier approach's well-established status, our research has identified significant inherent redundancies within the method. This finding suggests considerable scope for improvement, which is further elaborated in Section.\ref{F}.
	\par In recent years, the quantum Krylov-subspace method (QKSM)\cite{stair2020multireference,cortes2022quantum} has emerged as a valuable complement and, in certain scenarios, a preferable alternative to quantum phase estimation (QPE)\cite{abrams1999quantum,abrams1997simulation} for addressing the eigenpair problems. In this context, the QKSM is distinguished from the QPE in that it necessitates a multitude of shallow quantum circuits in contrast to the deeper circuits typically utilized by QPE. As a hybrid classical-quantum algorithms, the QKSM can effectively transform the original problem into a generalized eigenvalue problem of exponentially smaller dimension, which can be easily computed on a classical computer. Consequently, we immediately recognized the potential of the QKSM to reduce the redundancies of the Fourier approach. 
	\par Building upon these insights, we introduce the quantum Krylov-subspace method based linear solver(QKLS) which significantly diminishes the complexity factor $\kappa$ from polynomial to linear compared to the Fourier approach, thereby indirectly enhancing the HHL algorithm. Moreover, in the light of the findings from previous research\cite{stair2020multireference,cortes2022quantum} on the QKSM and our numerical experiment conducted on a 10-qubits non-trivial QLSP, our conclusions are rather moderate and conservative. This is because our analysis is predicated on the convergence studies of the classical Krylov-subspace method which is discussed in Section.\ref{3}.
	\par It is crucial to recognize that this enhancement comes with a trade-off. The implementation of the QKSM necessitates the use of a substantial number of shallow quantum circuits, that presents future researchers with considerations that demand careful evaluation. However, it is reasonable to anticipate that reducing the depth of quantum circuits will be essential even in the era of fault-tolerant quantum computers\cite{lin2022heisenberg}. This is due to the fact that overly deep circuits will introduce increased noise and a greater risk of decoherence.
	\par This paper is structured as follows: In Section.\ref{F}, we present our analysis demonstrating that there are substantial redundancies within the Fourier approach. In Section.\ref{2}, we apply the QKSM to the QLSP and propose a novel approach for evaluating the required elements, which is more efficient for the QKSM. In Section.\ref{3}, we present techniques for reconstructing quantum states based on the outcomes of the QKSM, along with the theoretical analysis of both query complexity and gate complexity. In Section.\ref{Numercial}, we exhibit numerical experiments and proceed with discussions.

	\section{Fourier Approach}\label{F}
	Childs et al.\cite{doi:10.1137/16M1087072} proposed the Fourier approach for QLSP. They provided an explicit analytical expression to demonstrate that within the domain $[-1,-1/\kappa] \cup [1/\kappa, 1]$, the inverse of $\mathcal{H}$ could be approximated with an error $\epsilon$ as follows:
	\begin{equation}\label{fa}
		\frac{1}{\mathcal{H}}\approx\frac{i}{\sqrt{2\pi}}\sum^{J-1}_{j=0}\Delta_y \sum^{k=K}_{k=-K}\Delta_z z_k e^{-{z_k}^{2}/2} e^{-i\mathcal{H}y_j z_k}
	\end{equation}
	where $y_{j} :=j \Delta_{y}$ , $z_{k} :=k \Delta_{z}$ , for some fixed $J ~=~ \Theta( \frac{\kappa} {\epsilon} \operatorname{l o g} ( \kappa/ \epsilon) )$ , $K ~=~ \Theta( \kappa\operatorname{l o g} ( \kappa/ \epsilon) )$ , 
	$\Delta_{y}=$  $\Theta( \epsilon/ \sqrt{\operatorname{l o g} ( \kappa/ \epsilon)} )$ , and $\Delta_{z}=\Theta( ( \kappa\sqrt{\operatorname{l o g} ( \kappa/ \epsilon)} )^{-1} )$ . 
	An explicit algorithm for linearly combining the terms in Eq.\eqref{fa} was also provided in Ref.\cite{doi:10.1137/16M1087072}, 
	thus the desired state $\ket{\tilde{\bm{x}}}$ could be prepared by mapping $\ket{\bm{b}}$ to $\ket{\tilde{\bm{x}}}$, where 
	$\ket{\Tilde{\bm{x}}}=\frac{i}{\sqrt{2\pi}}\sum^{J-1}_{j=0}\Delta_y \sum^{k=K}_{k=-K}\Delta_z z_k e^{-{z_k}^{2}/2} e^{-i\mathcal{H}y_j z_k}\ket{\bm{b}}$.
	\par The Fourier Approach introduced for solving the QLSP presented a novel framework that improved computational efficiency. Nonetheless, the following analysis reveals substantial potential for enhancement to the algorithm. Assume that $\mathcal{H}$ is invertible for the sake of argument, it becomes feasible to deduce the underlying essence of the expression delineated in  Eq.\eqref{fa} by summing and taking the union:
	\begin{align}\label{leak}
		\frac{1}{\mathcal{H}} &\approx \frac{i}{\sqrt{2\pi}} \sum_{j=0}^{J-1} \Delta_y \sum_{k=-K}^{K} \Delta_z z_k e^{-z_k^2/2} e^{-i\mathcal{H}y_j z_k} \nonumber \\
		&= \frac{i}{\sqrt{2\pi}} \sum_{j=0}^{J-1} \Delta_y \Big{(} \Delta_z^2 \sum_{k=1}^{K} k e^{-z_k^2/2} e^{-i\mathcal{H}y_j z_k}\nonumber \\
		&\qquad + \Delta_z^2 \sum_{k=1}^{K} (-k) e^{-z_k^2/2} e^{i\mathcal{H}y_j z_k} + \Delta_z^2 \cdot 0 \Big{)} \nonumber \\
		&= \frac{i}{\sqrt{2\pi}} \sum_{j=0}^{J-1} \Delta_y \Delta_z^2 \sum_{k=1}^{K} k e^{-z_k^2/2} (e^{-i\mathcal{H}y_j z_k} \nonumber \\
		& \qquad - e^{i\mathcal{H}y_j z_k}) \nonumber \\
		&= \frac{i}{\sqrt{2\pi}} \sum_{j=0}^{J-1} \Delta_y \Delta_z^2 \sum_{k=1}^{K} -2ki e^{-z_k^2/2} \sin(\mathcal{H} y_j z_k) \nonumber \\
		&= \frac{2}{\sqrt{2\pi}} \sum_{j=0}^{J-1} \Delta_y \Delta_z^2 \sum_{k=1}^{K} k e^{-z_k^2/2} \sin(\mathcal{H} y_j z_k). 
	\end{align}
	
	\par With Eq.\eqref{leak}, in Fourier approach, it is evidently that the representation above is expressed as a linear combination of the odd-order powers of that. 
	\par However, let the minimal polynomial of $\mathcal{H}$ be $f(\lambda) = \sum^{N}_{p=0} b_p \lambda^p$. According to the Hamilton-Cayley theorem, which states $f(\mathcal{H}) = 0$. The general inverse of $\mathcal{H}$ is given by $(-1/b_0)\sum^{N}_{p=1} b_p \mathcal{H}^{p-1}$, with the condition that $b_0 \neq 0$. This implies that the representation of $1/\mathcal{H}$ should contain both even-order powers of $\mathcal{H}$ and odd-order powers of $\mathcal{H}$. 
	Consequently, to achieve a desired state $\ket{\tilde{\bm{x}}}$, the Fourier approach  actually employs a substantial number of redundant terms to compensate for the absence of even-order powers of $\mathcal{H}$.

	\par
\begin{widetext}
		\begin{figure*}[htb]        
			\center{\includegraphics[width=\linewidth]  {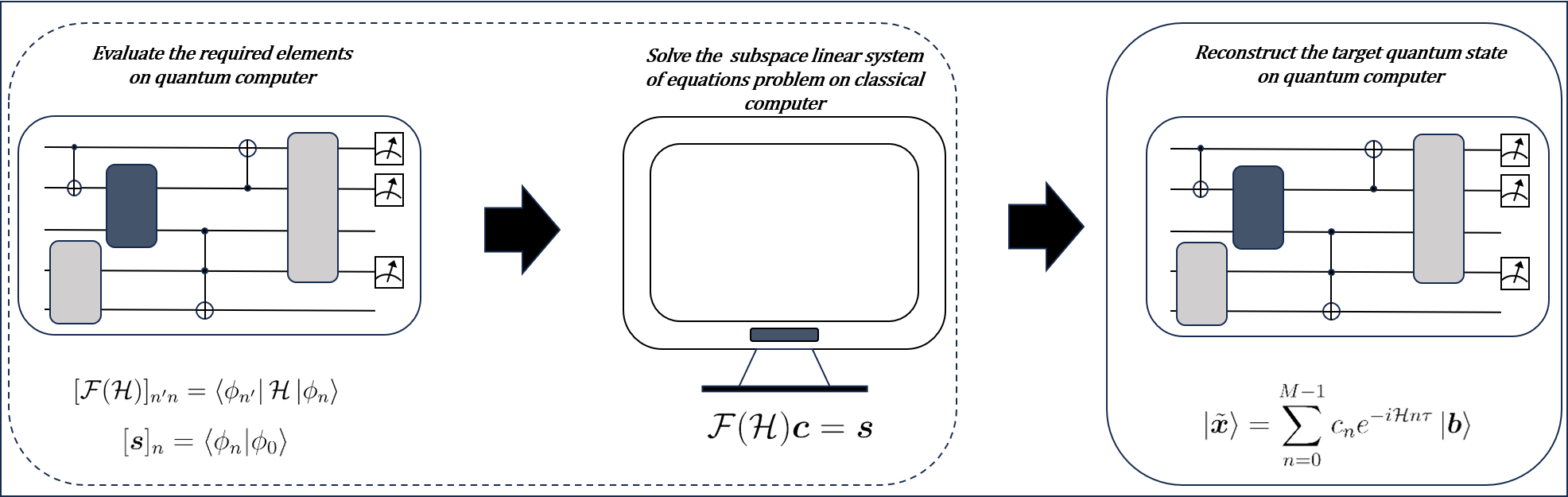}}  
			\caption{Overview of the quantum Krylov-subspace method based linear solver. The whole algorithm consists of three parts: 1. Evaluate each elements of $\mathcal{F}({\mathcal{H}})$ and $\bm{s}$ on the quantum computer; 2. Solve $\mathcal{F}({\mathcal{H}})\bm{c}=\bm{s}$ on the classical computer. The first two components of our algorithm constitute the QKSM; 3. Reconstruct $\ket{\Tilde{\bm{x}}}$ with the linear combinations of unitary operators on the quantum computer.}
			\label{skeme}
		\end{figure*}
	\end{widetext}

 In view of this, we introduce the quantum Krylov-subspace method based linear solver(as illustrated in Figure.\ref{skeme}). Our algorithm employ the QKSM as a subroutine, effectively reducing the redundancies of the Fourier method.
	\section{Quantum Krylov-subspace method}\label{2}
	\subsection{Apply the Quantum Krylov-subspace method to QLSP}\label{2.1}
	The QKSM was initially introduced by Stair et al.\cite{stair2020multireference}, who first associated real-time evolution operators with classical Krylov-subspace basis. This innovative approach culminated in the conception of the "quantum Krylov-subspace". The underlying concept of the QKSM is to represent the target state as a linear combination of a set of non-orthogonal quantum states, which are generated through real-time quantum dynamics. 
	\par Similar to the research approach detailed in  Ref.\cite{stair2020multireference} for eigenpair problems, we establish the connection between the classical Krylov-subspace method and the QKSM, and apply the QKSM to the QLSP.
	\par Classically, the approximate solution  $\widetilde{\ket{\bm{x}_M}}$ in the $M$ order Krylov-subspace $\mathcal{K}_M(\ket{\bm{b}},\mathcal{H})$, where $\mathcal{K}_M(\ket{\bm{b}},\mathcal{H}) := span\{\ket{\bm{b}},\mathcal{H}\ket{\bm{b}},\dots,\mathcal{H}^{M-1}\ket{\bm
		b}\}$, can be represented as a linear combination of the Krylov-subspace basis\cite{gutknecht2007brief}:
	\begin{equation} \label{classical krylov}
		\widetilde{\ket{\bm{x}_M}}=\sum_{k=0}^{M-1}\hat{c_k}\mathcal{H}^{k} \ket{\bm{b}}.
	\end{equation}
	\par Building upon the foundational principles of Krylov-subspace method, researchers have successfully developed a multitude of outstanding iterative algorithms\cite{hestenes1952methods,saad1986gmres,van1992bi,sonneveld1989cgs,paige1975solution}. However, performing Krylov-subspace method based iterative algorithms directly in quantum computers is not efficient. Nonetheless, a set of time-evolution operators can be employed to approximate $\mathcal{K}_M(\ket{\bm{b}},\mathcal{H})$. 
	
	Consider the set $\{e^{-i\mathcal{H}l\tau}\}_{l=0}^{M-1}$, where $\tau$ is the fixed time-step. Throughout the paper, the Planck constant is set to 1. A general state $\ket{\bm{x}_{M}}$ in $\mathcal{K}_M(\ket{\bm{b}},e^{-i\mathcal{H}\tau})$ can be represented as $\sum_{l=0}^{M-1} c_l e^{-i \mathcal{H}l \tau } \ket{\bm{b}}$. The exponential function is expanded into a Taylor series:
	\begin{eqnarray}    \label{eq}
		\ket{\bm{x}_{M}}&=&\sum_{l=0}^{M-1} c_l e^{-i \mathcal{H}l \tau } \ket{\bm{b}} \nonumber    \\
		~&=& \sum_{l=0}^{M-1}(c_l \sum_{k=0}^{M-1} \frac{(-i l \tau)^k}{k!})\mathcal{H}^k \ket{\bm{b}} + O(\tau ^M)\nonumber    
		\\
		~&=&\sum_{k=0}^{M-1}(\sum_{l=0}^{M-1} \frac{(-i l \tau)^k}{k!}c_l )\mathcal{H}^k  \ket{\bm{b}} + O(\tau ^M)\nonumber \\
		~&=&\sum_{k=0}^{M-1}\hat{c_k} \mathcal{H}^k  \ket{\bm{b}} +  O(\tau ^M) \approx \widetilde{\ket{\bm{x}_M}}.
	\end{eqnarray}
	\par In Light of the information, $\frac{(-i l \tau)^k}{k!}$ could be regard as an element of a square matrix $\bm{M}$, where $[\bm{M}]_{kl}:=\frac{(-i l \tau)^k}{k!}$, and $\bm{M}$ is a Vandermonde matrix. With Eq.\eqref{eq}, the coefficients $c_l$ can be selected to depict various linear combinations of the classical Krylov basis, ranging from $l=0\dots M-1$, up to higher-order terms.
	\par That is the origin of the QKSM heuristic\cite{stair2020multireference}. Establishing a conceptual bridge between the quantum and classical Krylov-subspace method enhances our comprehension of the QKSM's underlying principles. Nevertheless, this connection should not mislead one to believe that the two methods have equal capabilities. Experimental data from related research\cite{stair2020multireference,cortes2022quantum} indicates that the QKSM is not merely an extension of the classical Krylov-subspace method in the realm of quantum computation. Its performance significantly surpasses that of the classical Krylov-subspace method. 
	\par With Eq.\eqref{classical krylov} and Eq.\eqref{eq}, if $\widetilde{\ket{\bm{x}_M}}$ is $\epsilon$-approximate to $\bm{x}$, the approximate solution can be represented with quantum Krylov-subspace basis as:
	\begin{equation} \label{app}
		\ket{\Tilde{\bm{x}}}=\sum_{n=0}^{M-1} c_n e^{-i \mathcal{H}n \tau}\ket{\bm{b}}:=\sum_{n=0}^{M-1}c_{n} \ket{\phi_n}.
	\end{equation}
	
	\par Substituting Eq.\eqref{app} into the original problem and multiplying from the left by $\bra{\phi_{n^{\prime}}}$, we get:
	\begin{equation} \label{main}
		\mathcal{F}({\mathcal{H}})\bm{c}=\bm{s},
	\end{equation}
	where $\bm{c}=(c_{0},c_{1},\cdot\cdot\cdot,c_{M-1})^{T}$ is the column vector of expansion coefficients and the matrix  $\mathcal{F}({\mathcal{H}})$ and vector $\bm{s}$ defined as:
	\begin{equation} \label{matrix_elements}
		[\mathcal{F}({\mathcal{H}})]_{ n^{\prime} n}=\bra{\phi_ {n^{\prime}}}\mathcal{H}\ket{\phi_n} , [\bm{s}]_{ n^{\prime}} =\braket{\phi_ {n^{\prime}}|\phi_0}.
	\end{equation}
	\par Upon solving Eq.\eqref{main} on classical computer, the coefficients of Eq.\ref{app} can be readily derived. Prior to the solution process, we should evaluate the elements of $\mathcal{F}$ and $\bm{s}$.
	The elements of vector $\bm{s}$ can be efficiently evaluated with standard Hadamard-Test\cite{cleve1998quantum}. To evaluate the square matrix elements efficiently, we introduce a novel approach in Section.\ref{2.2}. Ultimately, the state $\ket{\Tilde{\bm{x}}}$ can be represented with the LCU on quantum computer and  we discuss this in Section.\ref{3}.
	\par \begin{remark}
		The application of the QKSM to the QLSP leads to the derivation of an approximation of the target state within the quantum Krylov-subspace. Based on the convergence analysis of the classical Krylov-subspace method \cite{shewchuk1994introduction}, the order $M$ required to obtain the desired $\ket{\Tilde{\bm{x}}}$ is of the order $\mathcal{O}(\kappa\log{\frac{1}{\epsilon}})$. This can be taken as an upper bound for the QKSM. Under the premise of achieving the same level of precision, the terms in Eq.\eqref{fa} are $\Theta(\frac{\kappa^2}{\epsilon}\log^2(\frac{\kappa}{\epsilon}))$. It is crucial to emphasize the marked decrease in redundancy associated with employing the QKSM compared to the Fourier approach. This reduction in redundancy optimizes the overall computational resources.
	\end{remark}
	\subsection{ A novel approach to evaluation the elements of $\mathcal{F}({\mathcal{H}})$ }\label{2.2}
	\par To evaluate the elements of the matrix $\mathcal{F}(\mathcal{H})$,  it is practical to categorize them into two types, the diagonal elements $\bra{\phi_n}\mathcal{H}\ket{\phi_n}$ and the off-diagonal elements$\bra{\phi_ {n^{\prime}}}\mathcal{H}\ket{\phi_n}$.   
	\par For the diagonal elements, the value can be determined by the expectation of the measurement of the observable $\mathcal{H}$. Typically, this expectation is calculated by decomposing $\mathcal{H}$ into Pauli operators and computing the value for each term. Thus, the number of these decomposition terms directly impacts the resource expenditure for evaluating diagonal elements. To mitigate this, Alonso-Linaje et al.\cite{alonso2021eva} proposed a quantum algorithm that approximates $\mathcal{H}$ using the time evolution operator $e^{-i\mathcal{H}t}$, thereby reducing the consumption resources associated with evaluating the diagonal elements. We omit a detailed description of the algorithm since our promotion of it will lead to a framework where the computation of the diagonal elements becomes a particular case of off-diagonal elements computation. Moreover, in certain cases,  alternative variants of the Hadamard-Test may also be effective\cite{yang2023phase,havlivcek2019supervised}. 
	\par For the off-diagonal elements, the situation is more intricate. Firstly, as $\ket{\phi_{n^{\prime}}}$ and $\ket{\phi_n}$ are distinct, we cannot determine the overlap by expectation of measurement of observable $\mathcal{H}$. Additionally, $\mathcal{H}$ is not guaranteed to be unitary, which means $\mathcal{H}$ cannot be directly implemented in quantum circuits. Therefore, we cannot apply the methods typically used for evaluating the diagonal elements to the off-diagonal elements. Inspired by the work of Alonso-Linaje et al.\cite{alonso2021eva}, we propose an extension to their algorithmic framework to address these problems.
	
	\par Consider the overlap between the quantum states $\ket{\phi_{n^{\prime}}}$ and $\ket{\phi_n}$ is given by the inner product $\bra{\phi_{n^{\prime}}}{e^{-i\mathcal{H}t}\ket{\phi_n}}$, where $\mathcal{H}$ is replaced by the time-evolution operator $e^{-i\mathcal{H}t}$. Expanding the exponential $e^{-i\mathcal{H}t}$ into a Taylor series:
	\begin{align}
		\label{tylor}
		\bra{\phi_{n^{\prime}}} e^{-i\mathcal{H}t} \ket{\phi_n}&=\bra{\phi_{n^{\prime}}} I \ket{\phi_n}+i \bra{\phi_{n^{\prime}}} \mathcal{H} \ket{\phi_n}t \nonumber \\
		&-\frac{ \bra{\phi_{n^{\prime}}} \mathcal{H}^{2} \ket{\phi_n} t^{2}}{2!}-\ldots 
	\end{align}
	\par Given the above Eq.\eqref{tylor}, we need to calculate the second term in right hand of Eq.(\eqref{tylor}). Upon rearranging the formula, the following expression is obtained:
	\begin{align}
		\bra{\phi_{n^{\prime}}} \mathcal{H} \ket{\phi_n}&=\frac{1}{it}(\bra{\phi_{n^{\prime}}} e^{-i\mathcal{H}t} \ket{\phi_n}-\bra{\phi_{n^{\prime}}} I \ket{\phi_n} \nonumber   \\
		&+\frac{ \bra{\phi_{n^{\prime}}} \mathcal{H}^{2} \ket{\phi_n} t^{2}}{2!}-\ldots) 
	\end{align}
	Taking a limit with respect to time $t$ on this result, we could obtain:
	\begin{align} \label{overlap}
		\bra{\phi_{n^{\prime}}} \mathcal{H} \ket{\phi_n}&=\frac{1}{it}(\bra{\phi_{n^{\prime}}} e^{-i\mathcal{H}t} \ket{\phi_n}-\bra{\phi_{n^{\prime}}} I \ket{\phi_n}) \nonumber \\
		& + O(t).
	\end{align}
	\par It should be noticed that both terms in right hand of Eq.\eqref{overlap} can be efficiently determined with the standard Hadamard-Test. This is reasonable, as $e^{-i\mathcal{H}t} $ and all the other $e^{-i\mathcal{H}k\tau}$ are commutative. Then, we could find that:
	\begin{align}
		Re{ \bra{\phi_{n^{\prime}}} \mathcal{H} \ket{\phi_n}}&\approx \frac{1}{t}\Big{(}Im{\bra{\phi_{n^{\prime}}} e^{-i\mathcal{H}t} \ket{\phi_n}}\nonumber \\
		&-Im{\bra{\phi_{n^{\prime}}} I \ket{\phi_n}}\Big{)}
	\end{align}
	\begin{align}
		Im{ \bra{\phi_{n^{\prime}}} \mathcal{H} \ket{\phi_n}}&\approx \frac{1}{t}\Big{(}Re{\bra{\phi_{n^{\prime}}} I \ket{\phi_n}} \nonumber \\
		&-Re{\bra{\phi_{n^{\prime}}} e^{-i\mathcal{H}t} \ket{\phi_n}}\Big{)}
	\end{align}
	\par
\begin{figure}[!h]        
		\centering
		\includegraphics[width=\linewidth]  {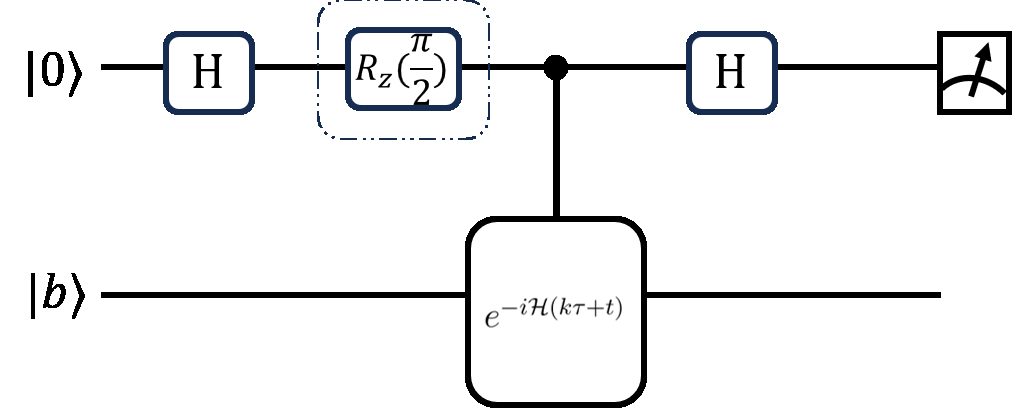}  
		\caption{Quantum circuit for implementing the variation of the novel approach. The circuit shown here is the computation of a general elements $\bra{\phi_{n^{\prime}}} \mathcal{H} \ket{\phi_n}$, where $k$=$n-n^{\prime}$. The operator H is Hadamard gate. The operator inside the dashed box is activated when one is going to evaluate the imaginary part of the overlap.}
		\label{hadamard}
	\end{figure}

 After some classical post-processing, $\bra{\phi_{n^{\prime}}} \mathcal{H} \ket{\phi_n}$ can be determined. The circuit for implementing the novel approach is illustrated in Figure.\ref{hadamard}.
	\section{Reconstruction of the target state}\label{3}
	\par The last step of our algorithm is to reconstruct the target state $\ket{\Tilde{\bm{x}}}$ with the LCU on the quantum computers. This process is similar to the Fourier approach, and we have not generalized or improved it in the present work. To ensure the integrity of the QKLS, we would elaborate the LCU and give the analysis of computational complexities based on Ref.\cite{doi:10.1137/16M1087072}. Additionally, the complexity analysis in Ref.\cite{doi:10.1137/16M1087072} was predicated upon a Hamiltonian simulation algorithm, which is a query model\cite{berry2015simulating}, we shall give the query complexity and gate complexity (2-qubit gate) respectively.  
	\subsection{Overview of the linear combinations of unitary operators }\label{3.1}
	\par We employ the LCU to implement the quantum operation $\mathcal{M}=\sum_{n=0}^{M-1} c_n e^{-i \mathcal{H}n \tau}$ via a quantum circuit. $\mathcal{M}$ is designed to map $\ket{\bm{b}}$ to $\ket{\Tilde{\bm{x}}}$. In accordance with Ref.\cite{doi:10.1137/16M1087072}, we assume that $c_n > 0$ for all $n$ and $\log(M-1)=m$ is an integer. With the lemma.6 and lemma.7 in Ref.\cite{doi:10.1137/16M1087072}, we could implement $\mathcal{M}$ as following:
	\par Let $\mathcal{V}$ be an operator acts on the first register and satisfies $\mathcal{V}\ket{\bm{0}^m}=\frac{1}{\sqrt{c}}\sum_{n=0}^{M-1}\sqrt{c_n}\ket{n}$, where $c:=\sum_{n=0}^{M-1} c_n$. Let $\mathcal{U}:=\sum_{n=0}^{M-1}\ket{n}\bra{n}\otimes e^{-i\mathcal{H}n\tau }$ acts on the whole circuit. Then the operator $\mathcal{W}:=\mathcal{V}^{\dagger}\mathcal{U}\mathcal{V}$ satisfies :
	\begin{equation}
		\mathcal{V}^{\dagger}\mathcal{U}\mathcal{V}\ket{\bm{0}^m}\ket{\bm{b}}=\frac{1}{c}\ket{\bm{0}^m}\mathcal{M}\ket{\bm{b}}+\ket{\bm{\Psi}^{\bot}},
	\end{equation}
		\begin{figure}[htb]        
		\centering
		\includegraphics[width=\linewidth]  {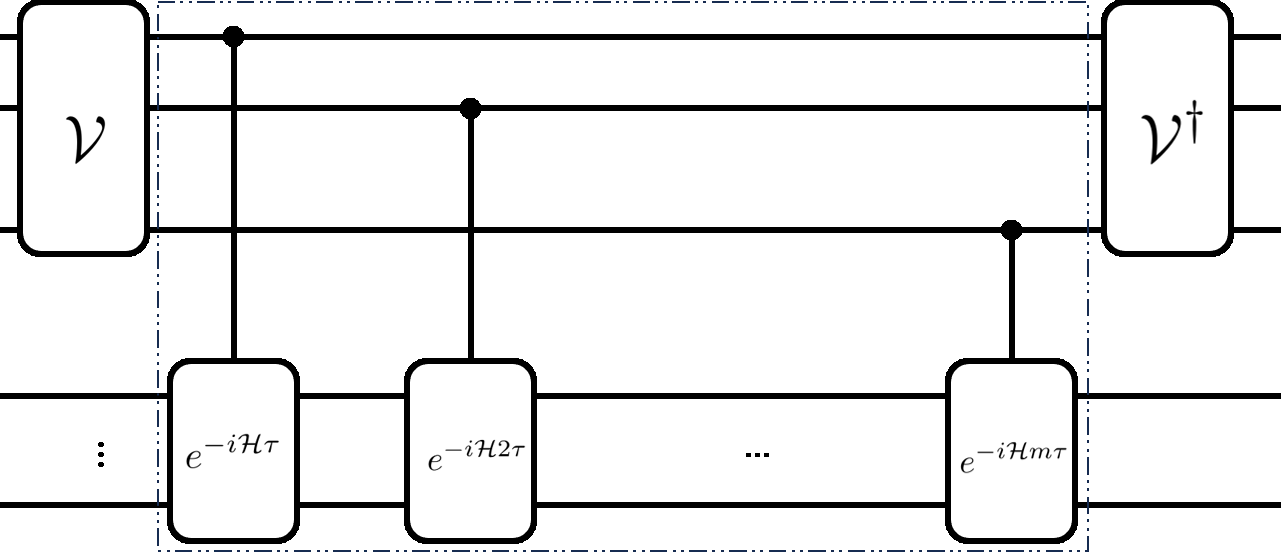}  
		\caption{The main part $\mathcal{W}$ for implementing linear combination of unitary operators. 
			The part in the dashed box is the explicit circuit of $\mathcal{U}$, which is consist of the set
			$\{c\mhyphen e^{-i\mathcal{H}2^{r}\tau}\}_{r=0}^{m}$.}
		\label{LCU}
	\end{figure}
	where $(\ket{\bm{0}}\bra{\bm{0}}\otimes\mathcal{I})\ket{\bm{\Psi^{\bot}}}=0$. Upon measuring the first m-qubits of the state and obtaining the outcome $\ket{\bm{0}^m}$, the state of the second register is proportionally represented as $\mathcal{M}\ket{\bm{b}}$ with a probability given by $(\Vert \mathcal{M}\ket{\bm{b}}\Vert/c)^2$. Figure.\ref{LCU} illustrates the explicit quantum circuit designed for the implementation of $\mathcal{W}$.

	\par In this scenario, given one capability to replicate the input state $\ket{\bm{b}}$ with $\mathcal{P}$, the amplitude amplification\cite{brassard2002quantum} could be utilized to construct the state with a high probability. Leveraging the quadratic speedup facilitated by amplitude amplification, the desired state could be determined with $\mathcal{O}(c/\Vert \mathcal{M}\ket{\bm{b}}\Vert)$ repetition. With Corollary.10 in Ref.\cite{doi:10.1137/16M1087072}, the norm $\Vert \mathcal{M}\ket{\bm{b}}\Vert$ is found to be greater than or equal to 1 when the eigenvalues of $\mathcal{H}$ fall within the interval $[-1,-1/\kappa]\cup[1/\kappa, 1]$. Consequently, the desired state could be determined with making $\Theta(c)$ uses of $\mathcal{P}$, $\mathcal{U}$, and $\mathcal{V}$ in expectation.
	
	\subsection{Query complexity}\label{3.2}
	The query complexity of the LCU is determined by the query complexity of $\mathcal{U}$, as detailed in Section.\ref{3.1}. This is because the implementation 
	of $\mathcal{V}$ query-free. The total query complexity for the LCU, which is consistent with the total query complexity of $\mathcal{U}$, can be calculated by considering two main factors: the query complexity associated with each individual simulation and the number of Hamiltonian simulation instances, whih is $\Theta(c)$. 
	\par From Section.\ref{2.1}, we could know that the longest evolution times is $\tau_{max}=(M-1)\tau=\mathcal{O}(\kappa\log{\frac{1}{\epsilon}})$, where $\tau<1$. $\tau_{max}$ can be regarded as a upper bound of the evolution time.
	\par The cumulative error in the procedure is guaranteed not to exceed $\epsilon$, provided each individual simulation's error is bounded by the average error $\epsilon' = \mathcal{O}(\epsilon/c)$. Moreover, the evaluation of $c_n$ is performed on classical computers. Before performing LCU, we can always scale $c=\mathcal{O}\left(1\right)$. Hence, we obtain $\epsilon^{\prime}=\mathcal{O}(\epsilon)$ 
	\par Using the Hamiltonian simulation algorithm in Ref.\cite{Berry_2015}, the query complexity of simulating a $d$-sparse Hamiltonian for time $t$ with error at most $\epsilon^{\prime}$ is 
	\begin{equation}
		\mathcal{O}\big(d\Vert\mathcal{H}\Vert_{max} t\log\left(\Vert\mathcal{H}\Vert t/\epsilon^{\prime}\right)\big)=\mathcal{O}\big(d t\log(t/\epsilon^{\prime})\big),
	\end{equation} 
	where we drop the denominator and use the assumption that $\Vert\mathcal{H}\Vert_{max}\leq\Vert\mathcal{H}\Vert\leq1$. 
	\par We can use $\tau_{max}$ as a upper bound for all $t\leq \tau_{max}$. That is the query complexity of each individual simulation can be upper bounded by $\mathcal{O}\big(d\tau_{max}\log(\tau_{max}/\epsilon^{\prime})\big)$.
	Then the total query complexity of LCU equals the number of Hamiltonian simulation instances multiplied by the upper bounded query complexity of each individual simulation: 
	\begin{small}
		\begin{align}
			\mathcal{O}\big(cd\tau_{max}\log(\tau_{max}/\epsilon^{\prime})\big)=\mathcal{O}\big(d\kappa\log(1/\epsilon)\log(\kappa/\epsilon)\big).
		\end{align}
	\end{small}
	
	\subsection{Gate complexity}\label{3.3}
	To analyze the gate complexity of LCU, we need to consider the gate complexity of $\mathcal{V}$ and $\mathcal{U}$ defined in Section.\ref{3.1} respectively. 
	\par Firstly, we could utilize $\mathcal{O}(M)$ gates\cite{shende2005synthesis} to implement the operator $\mathcal{V}$, which maps m-qubits vacuum state $\ket{0^m}$ to $\frac{1}{\sqrt{c}}\sum_{k=0}^{M-1}\sqrt{c_k}\ket{k}$. 
	\par Then we consider the operator $\mathcal{U}$. The explicit circuit of operator $\mathcal{U}$ is to implement sequentially the set $\{c\mhyphen e^{-i\mathcal{H}2^{r}\tau}\}_{r=0}^{m}$, where $c\mhyphen e^{-i\mathcal{H}2^{r}\tau}$ is the controlled version of $e^{-i\mathcal{H}2^{r}\tau}$. Therefore, we could sum the gate cost of $c\mhyphen e^{-i\mathcal{H}2^r\tau}$ from $r=0$ to $r=m =\mathcal{O}\big(\log(\kappa\log\frac{1}{\epsilon})\big)$ to get the total gate cost of $\mathcal{U}$. Additionally, for a given unitary operator $U$ with gate complexity $G$, the gate complexity of $c\mhyphen U$ is $\mathcal{O}(G)$.  
	\par From Ref.\cite{Berry_2015,doi:10.1137/16M1087072}, we could know that the gate complexity of simulating a $d\mhyphen sparse$ Hamiltonian $\mathcal{H}$ for time $t$ with error $\bar{\epsilon}$, where $t\leq1$, is
	\begin{equation} \label{hamiltonian simulation}
		\mathcal{O}\Big(\big(dt+1 \big)\big(\log N + {\log}^{2.5}( t/\bar{\epsilon})\big)\big(\log( t/\bar{\epsilon})\big)\Big),
	\end{equation}
	where we drop the denominator and use the assumption that $\Vert\mathcal{H}\Vert_{max}\leq\Vert\mathcal{H}\Vert\leq1$.
		
	\par The error in the implementing of $\mathcal{U}$ is guaranteed not to exceed ${\epsilon}^{\prime}$, provided the error to implement the operator $e^{-i\mathcal{H}2^r\tau}$ is bounded by $\bar{\epsilon}={\mathcal{O}}(\epsilon^{\prime}/m)=\mathcal{O}\big(\epsilon^{\prime}/\log(\kappa\log\frac{1}{\epsilon})\big)$.
	\begin{widetext}
		\begin{table*}[htb]
			\caption{Comparison between the Fourier approach in \cite{doi:10.1137/16M1087072} 
				and the quantum Krylov-subspace method based linear solver(QKLS).}\label{table1}
			\scriptsize
			\begin{tabular*}{\hsize}{@{}@{\extracolsep{\fill}}ccc@{}}
				
				\toprule
				$Algorithm $  & $Query Complexity$ & $Gate Complexity$ \\
				\midrule
				\midrule
				$FA$  &{ $\mathcal{O}\big(d\kappa^{2}\log^{2.5}(1/\epsilon)\big)$}   &{$\mathcal{O}\Big(\big(d\kappa^{2}\log^{2.5}{\frac{\kappa}{\epsilon}}\big)\big(\log{N}+\log^{2.5}\frac{\kappa}{\epsilon}\big)\Big)$}\\
				\midrule
				$QKLS$ &{ $\mathcal{O}\big(d\kappa\log(1/\epsilon)\log(\kappa/\epsilon)\big)$} &{ $\mathcal{O}\Big(\big(d\kappa\log^{2}{\frac{1}{\epsilon}}\big)\big(\log{N}+\log^{2.5}\frac{1}{\epsilon}\big)\Big)$} \\
				\bottomrule
			\end{tabular*}
		\end{table*} 
	\end{widetext}
	\par The longest time of Hamiltonian simulation of $\mathcal{U}$ is $\tau_{max}$. Then the cost of simulation $\mathcal{H}$ for any time $t \leq \tau_{max}$, to error $\bar \epsilon$ can be bound by
	\begin{equation}\label{upperbound}
		\mathcal{O}\Big(\big(dt+1\big)\big(\log{N}+\log^{2.5}({\tau_{max}/\bar{\epsilon}})\big)\log\big({\tau_{max}/\bar{\epsilon}}\big)\Big). 
	\end{equation}
	That means the total cost of implementing $\mathcal{U}$ is the sum of Eq.\eqref{upperbound} over $t$, which is
	\begin{small}
		\begin{align}
			\mathcal{O}\Big(\big(d\kappa\log{\frac{1}{\epsilon}}\big)\big(\log{N}+\log^{2.5}({\tau_{max}/\bar{\epsilon}})\big)\log\big({\tau_{max}/\bar{\epsilon}}\big)\Big).
		\end{align}
	\end{small}
	\par The overall gate complexity of LCU equals the number of Hamiltonian simulation instances multiplied by the gate complexity of $\mathcal{U}$, since the gate complexity of $\mathcal{V}$ is dominated by that of $\mathcal{U}$. Invoking Hamiltonian simulation $\Theta(c)$ times, then the overall gate complexity of LCU is: 
	\begin{equation}\label{gate_complexity}
		\mathcal{O}\Big(\big(d\kappa\log^{2}{\frac{1}{\epsilon}}\big)\big(\log{N}+\log^{2.5}\frac{1}{\epsilon}\big)\Big),
	\end{equation}
	where we substitute the values of $\tau_{max}$, $\bar{\epsilon}$ and using $\epsilon^{\prime}=\mathcal{O}(\epsilon/c)=\mathcal{O}(\epsilon)$. 
	\par We shall not delve further into the complexity analysis within Section.\ref{2.2} in this present work, as we posit it is dominated by LCU. We list the computational complexities of both QKLS and the Fourier approach in Table.\ref{table1}. It is not difficult to find that our algorithm improves the dependence on both $\kappa$ and $\epsilon$ with both additionally classical and quantum preprocessing.
\begin{figure}[t]
	\centering
	\subfloat[]{\label{4a}\includegraphics[width=.48\textwidth,height=4.5cm]{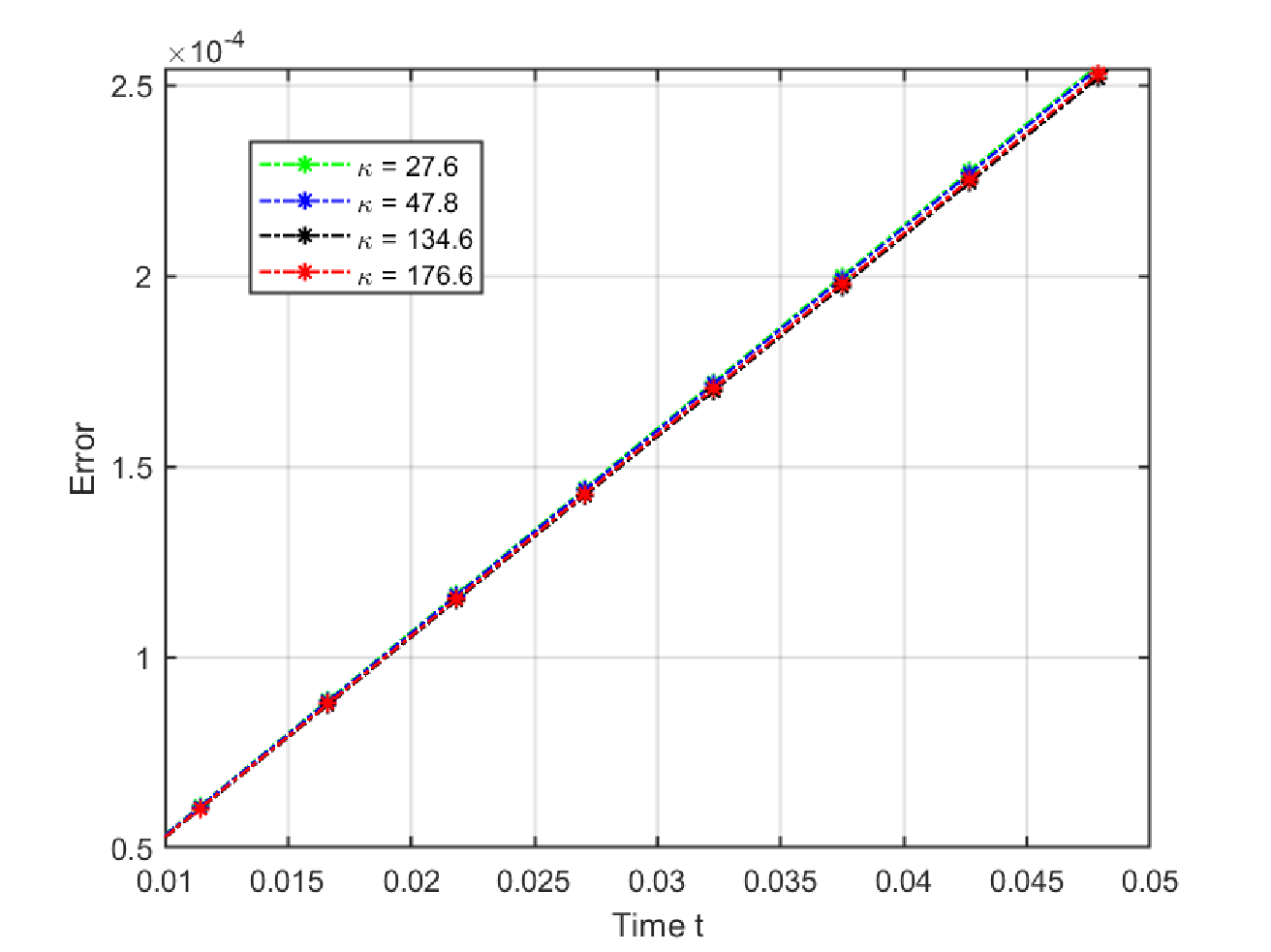}}\\
	\subfloat[]{\label{4b}\includegraphics[width=.48\textwidth,height=4.5cm]{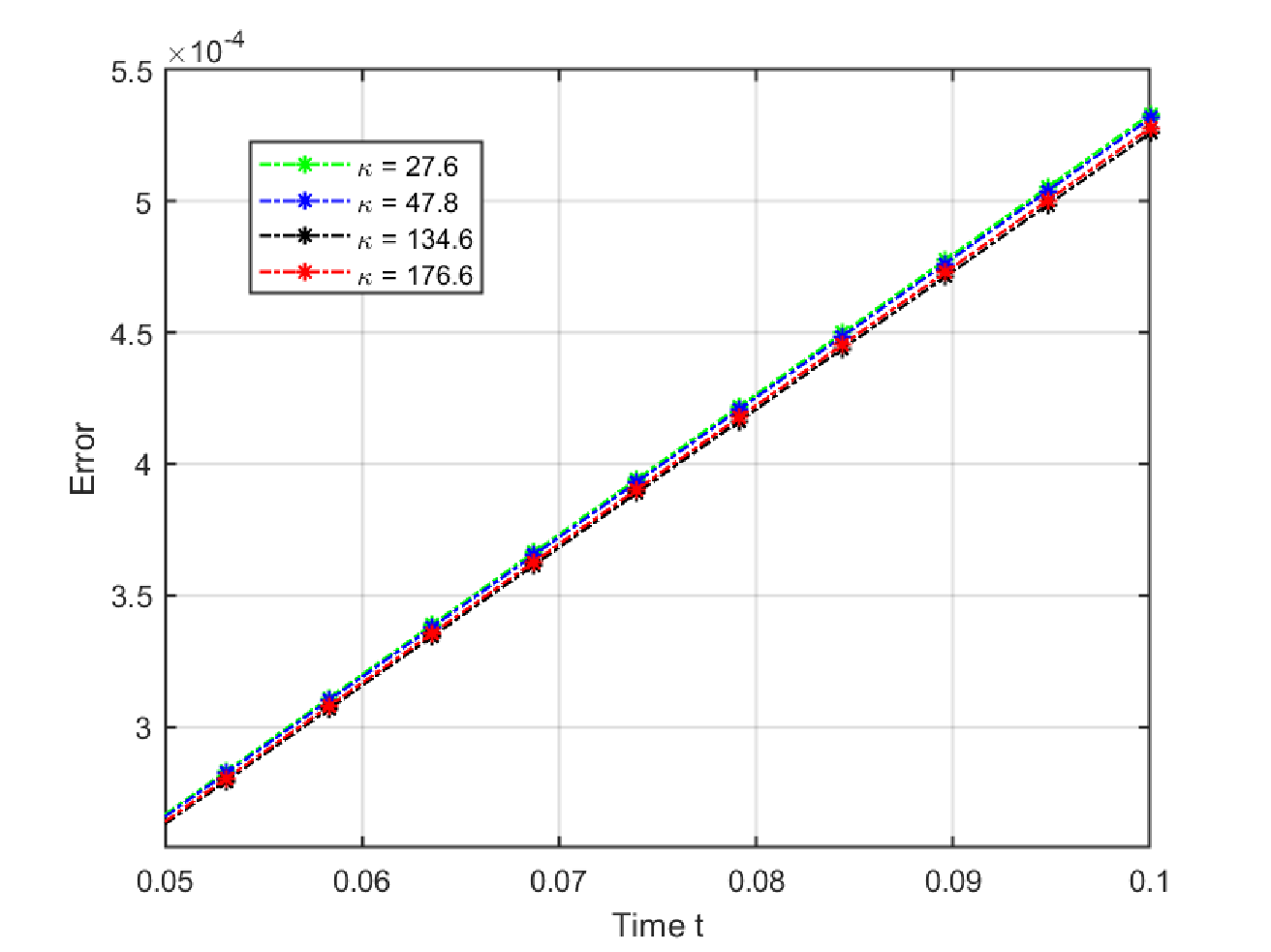}}
	\caption{This numerical experiment of the novel approach introduced in Section.\ref{2.2}, presenting curves for $\tau$ set to 0.1 and various $\kappa$ values: 27.6, 47.6, 134.6, and 176.6. It is important to recognize that Figures \ref{4a} and \ref{4b} together encompass the complete sampling time $t$ over the interval (0.01, 0.1).}
	\label{hadamard_num}
\end{figure}
	
	\section{Numerical experiment}\label{Numercial}
	
	To demonstrate the performance of the quantum Krylov-subspace method based linear solver in solving QLSP, we conducted numerical simulations on the nontrivial sparse matrix as defined in Ref.\cite{He_2017}:
	\begin{equation}\label{example}
		\mathcal{H}=\frac{1}{\zeta}\left( \sum_{j=1}^{n} X_{j}+J \sum_{j=1}^{n-1} Z_{j} Z_{j+1}+\eta\mathcal{I}\right)
	\end{equation}
	\begin{equation}
		\ket{\bm{b}}=H^{\otimes n}\ket{\bm{0}},
	\end{equation}
	where $X_{j}$, $Z_{j}$ in Eq.\ref{example} is Pauli operator $\sigma_x$, $\sigma_z$ acts on the $j\mhyphen th$ qubit respectively.
	We assign the value of $J$ to be $0.1$ and select distinct values for $\eta$ and $\zeta$ to construct matrices with varying condition numbers. The quantum state $\ket{\bm{b}}$ is prepared by applying a sequence of Hadamard gates to all qubits, making $\ket{\bm{b}}$ a non-trivial state for the QLSP.

	\par Firstly, we calculated the overlaps with the novel approach introduced in Section.\ref{2.2} using numerical simulations on the Pennylane \cite{bergholm2022pennylane} platform and then benchmarked it against the actual 
	\clearpage
	\begin{widetext}
		\begin{figure*}[t!]
			\centering
			\subfloat[]{\label{a}
				\includegraphics[width=0.4\linewidth,height=4cm]{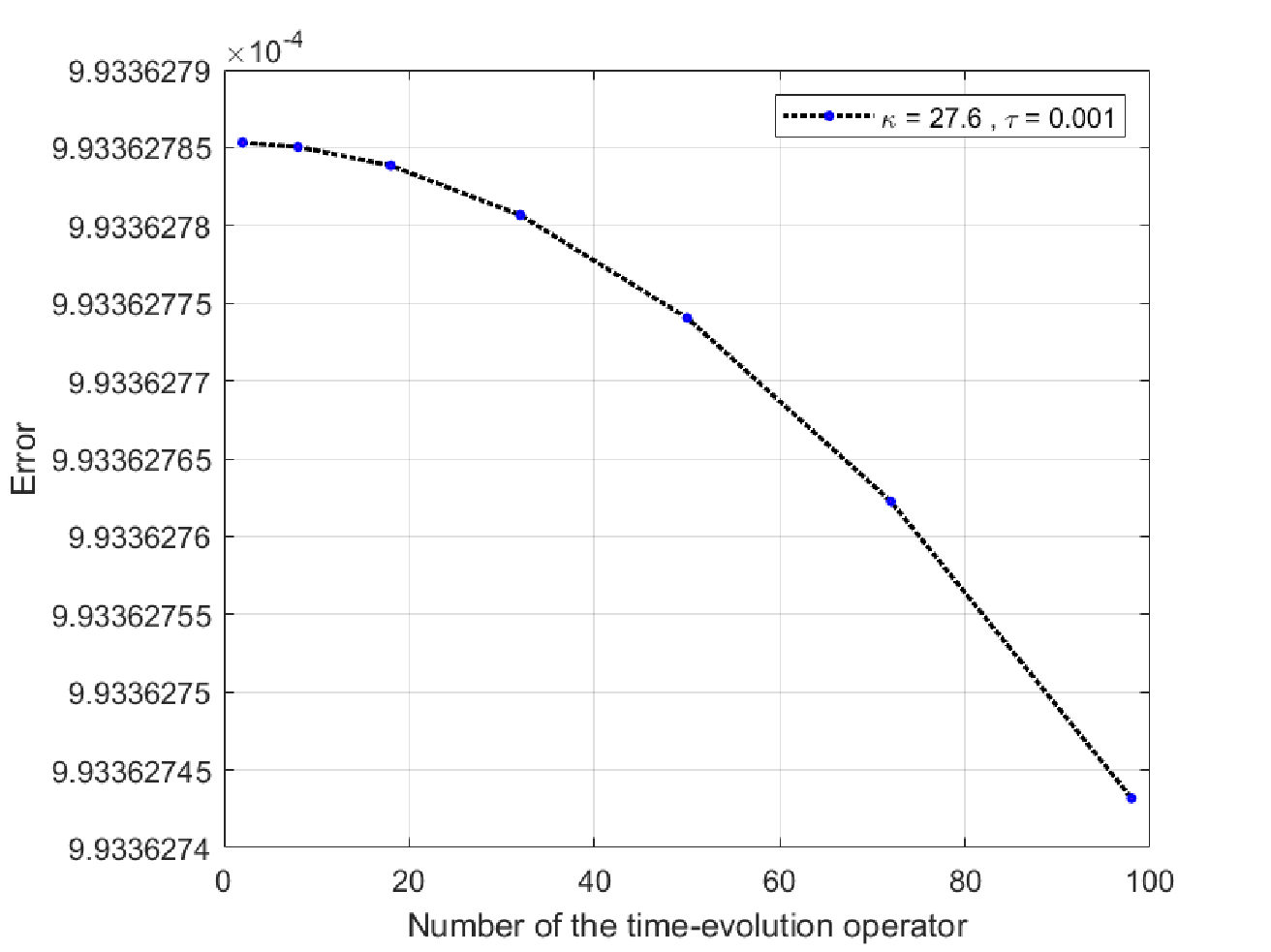}
			}\hfill
			\subfloat[]{\label{b}
				\includegraphics[width=0.4\linewidth,height=4cm]{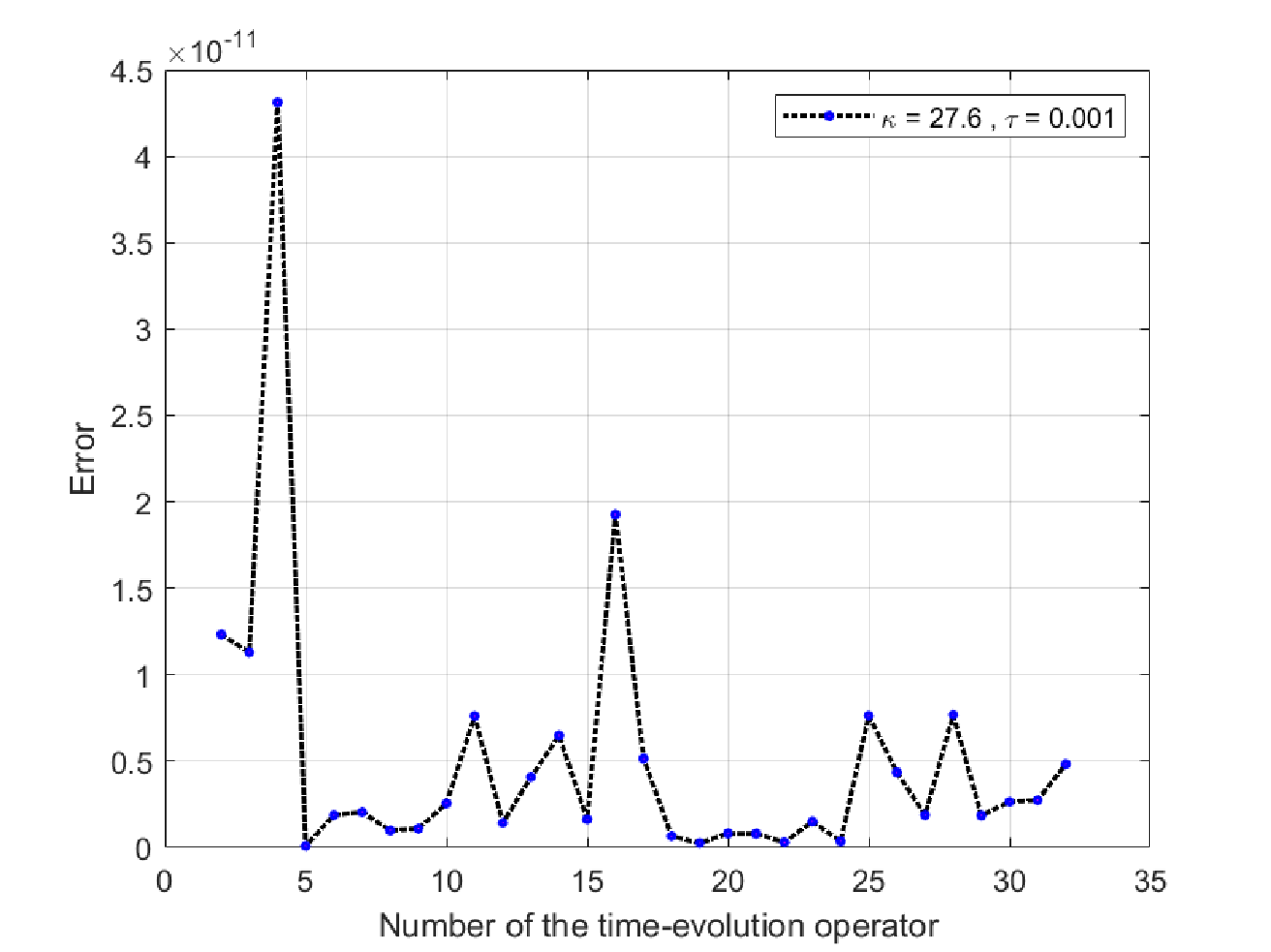}
			}\\
			\subfloat[]{\label{c}
				\includegraphics[width=0.4\linewidth,height=4cm]{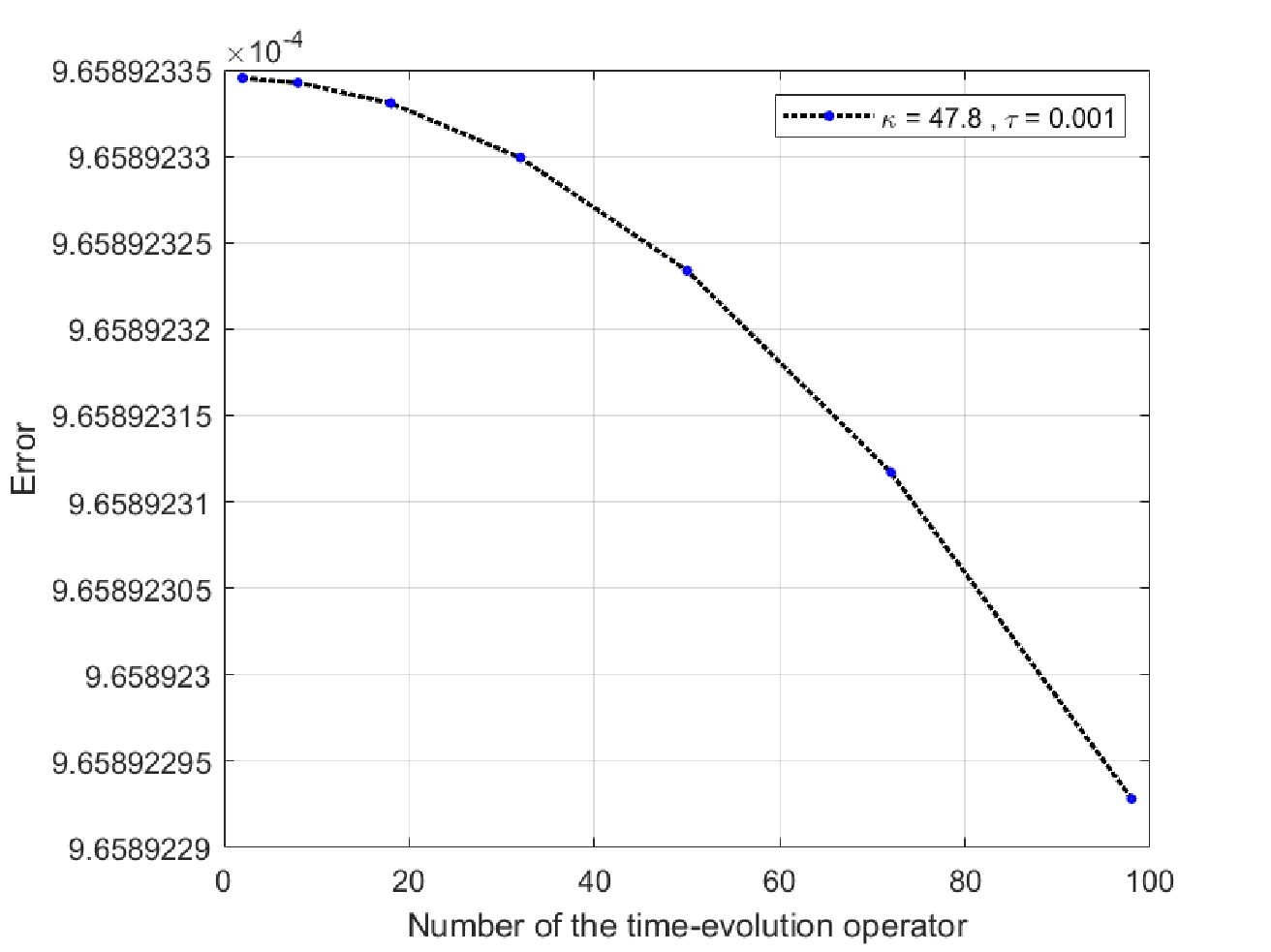}
			}\hfill
			\subfloat[]{\label{d}
				\includegraphics[width=0.4\linewidth,height=4cm]{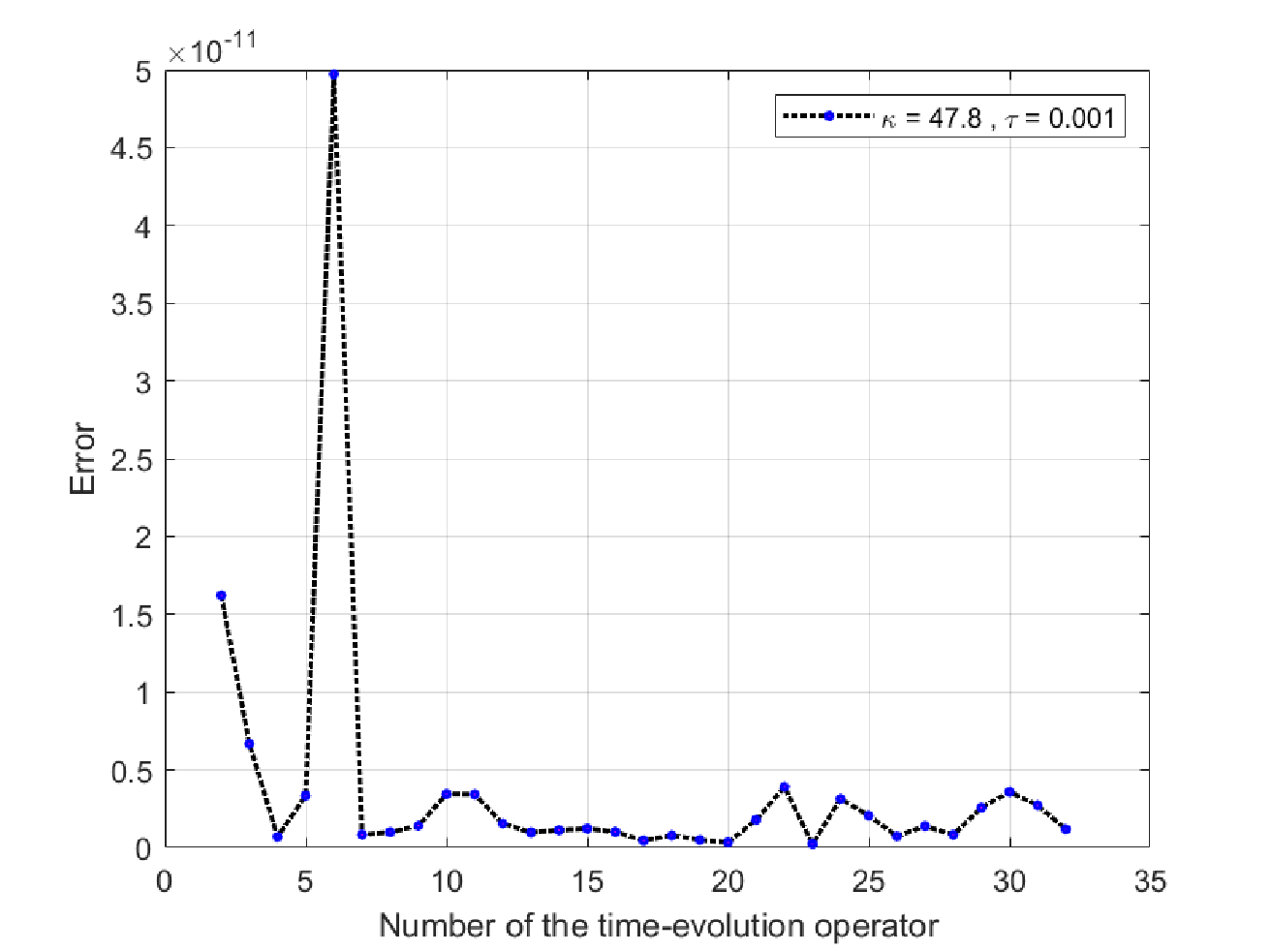}
			}\\
			\subfloat[]{\label{e}
				\includegraphics[width=0.4\linewidth,height=4cm]{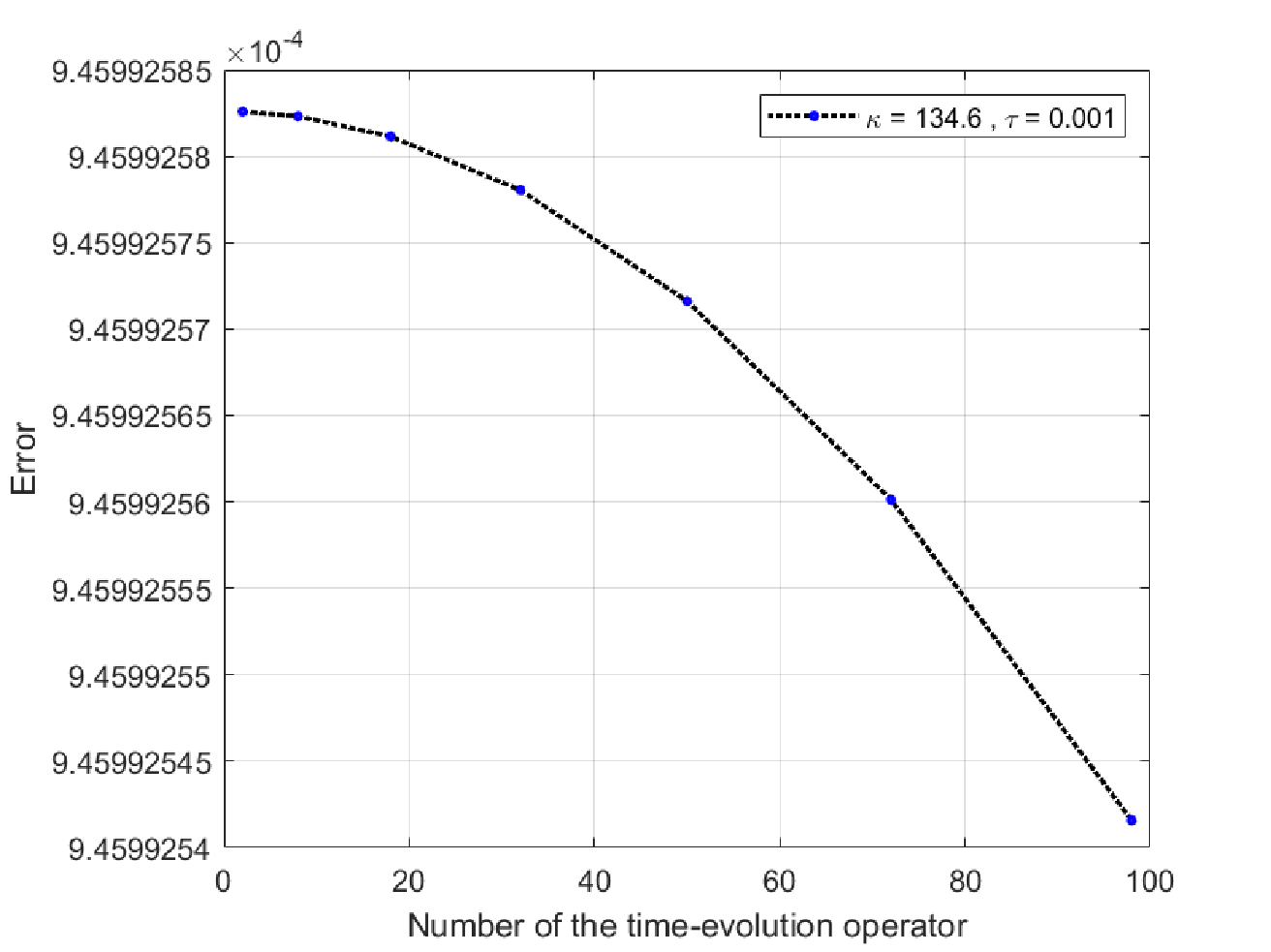}
			}\hfill
			\subfloat[]{\label{f}
				\includegraphics[width=0.4\linewidth,height=4cm]{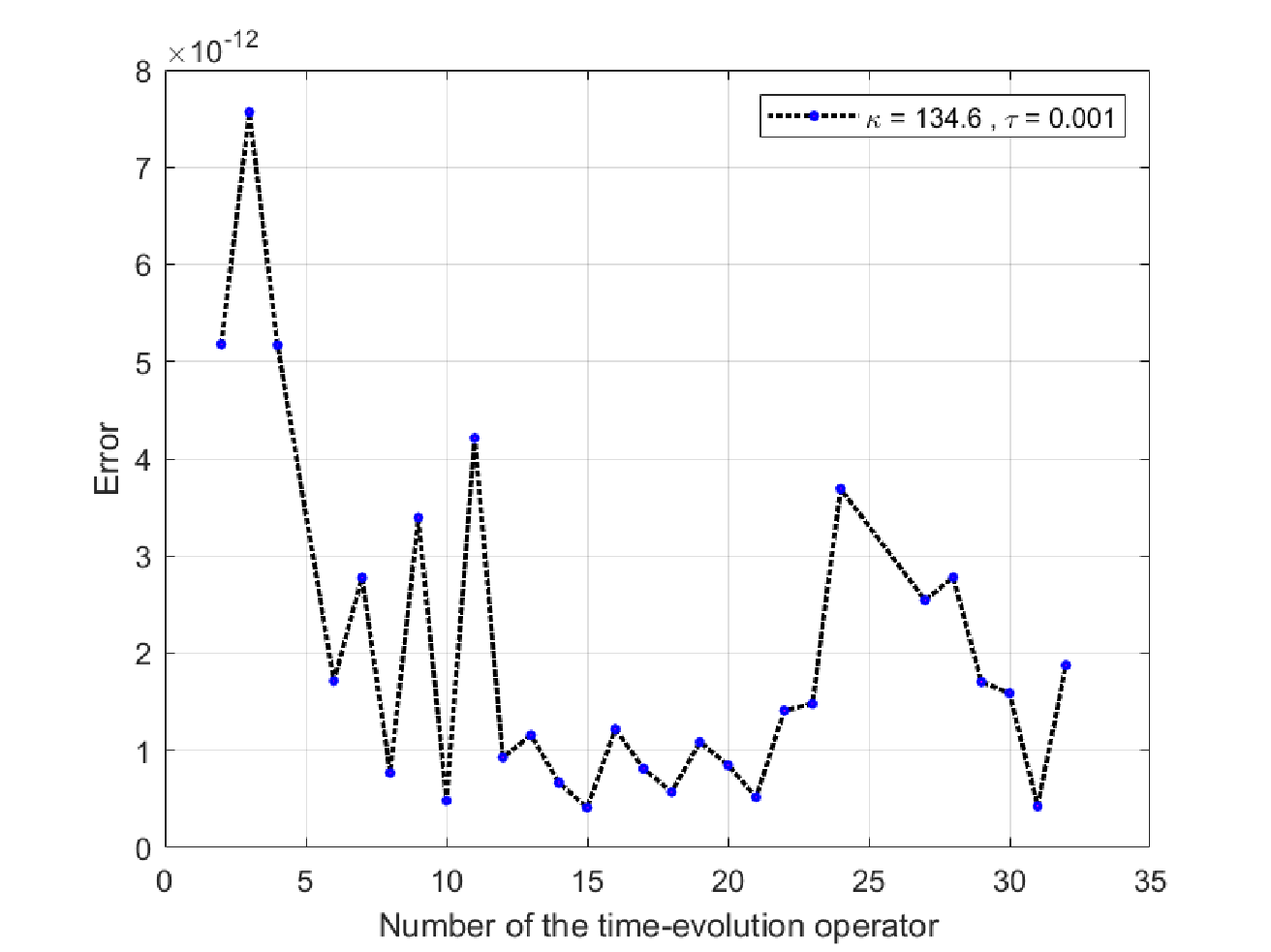}
			}\\
			\subfloat[]{\label{g}
				\includegraphics[width=0.4\linewidth,height=4cm]{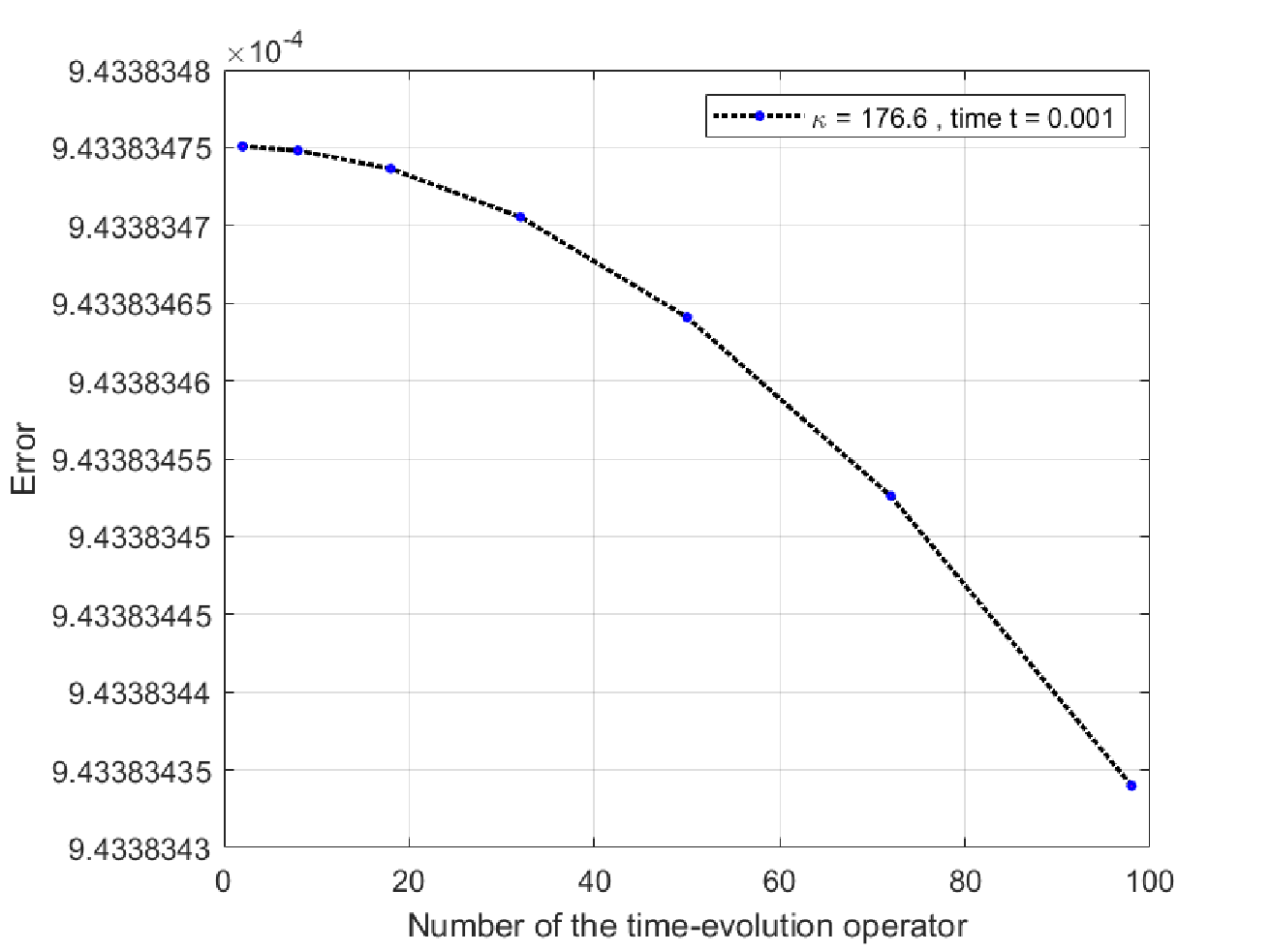}
			}\hfill
			\subfloat[]{\label{h}
				\includegraphics[width=0.4\linewidth,height=4cm]{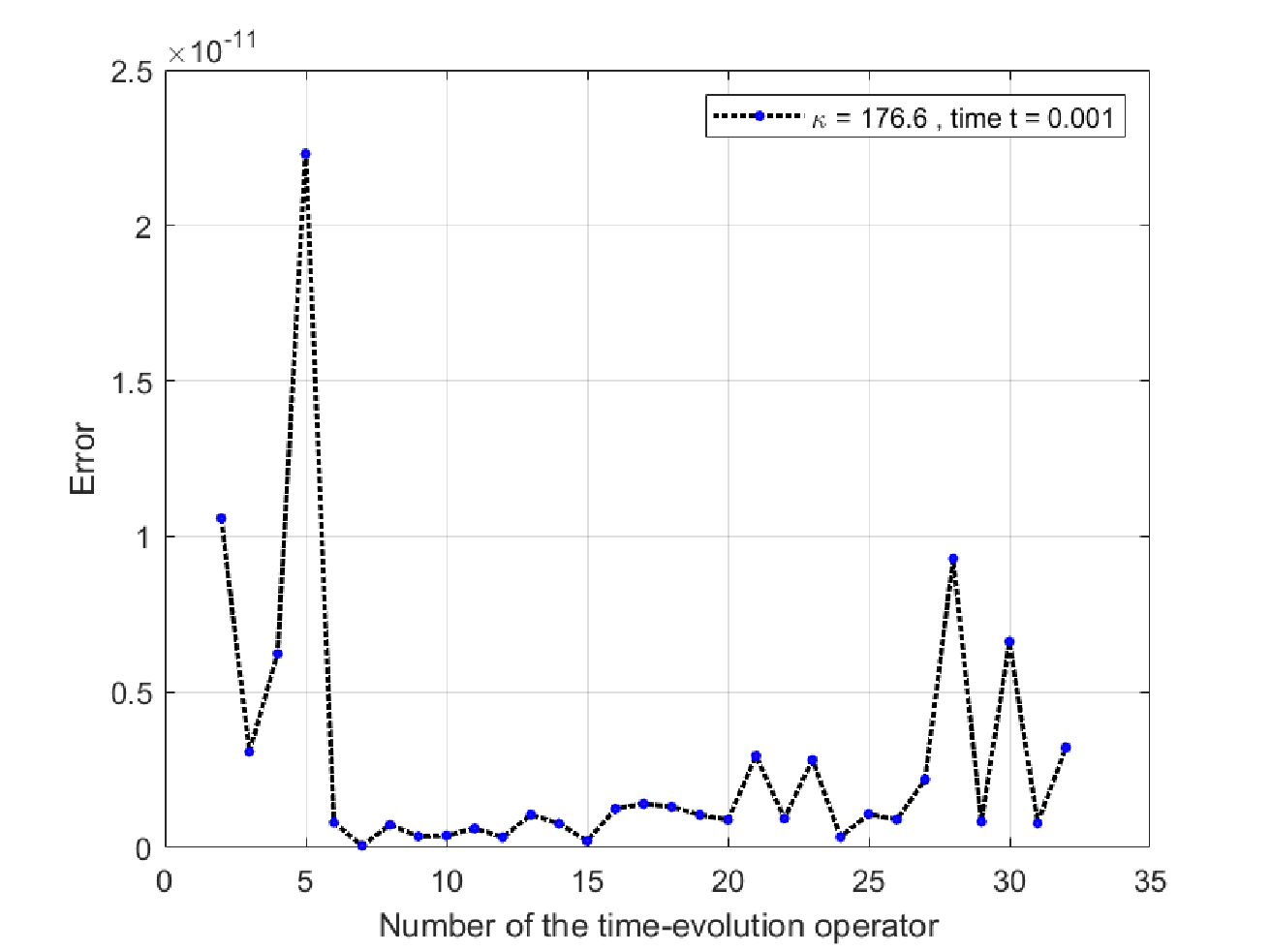}
			}  
			\caption{This experiment numerically compares the performance of the quantum Krylov-subspace method based linear solver with the Fourier approach using a 10-qubit system. Figures \ref{b}, \ref{d}, \ref{f} and \ref{h}, with QKLS, display curves for a fixed time step $\tau$ of 0.001, with different condition numbers $\kappa$ set to 27.6, 47.6, 134.6, and 176.6 respectively. Similarly, Figures \ref{a}, \ref{c}, \ref{e} and \ref{g} present curves for the same values of $\tau$ and $\kappa$ with the Fourier approach.}
			\label{QKSMNM}
		\end{figure*}
	\end{widetext}
	value. In our numerical experiments, we approximated the real-time evolution using the Trotter-Suzuki decomposition\cite{suzuki1977convergence}. The error metric employed in this comparison was the absolute distance between the two values, i.e. $\lVert a - b \rVert$. Figure.\ref{hadamard_num} suggested that the error tended to rise linearly with an increase in time $t$. This observed scaling is consistent with the analysis presented in Section.\ref{2.2}.
	\par  We proceeded to simulate the entire algorithm classically, which means that both our algorithm and the Fourier approach were operating under ideal conditions. In this context, we employed the error metric given by the distance between the norm of the inner product between vectors $\Vec{a}$ and $\Vec{b}$ and $1$, i.e. $\lVert 1- \langle\Vec{a}, \Vec{b}\rangle \rVert$.  
	
	Figure.\ref{QKSMNM} indicated that by using fewer time-evolution operators $e^{-i\mathcal{H}k\tau}$, our algorithm could achieve much better approximations in comparison to the Fourier method. These findings supported the analysis in both Section.\ref{F} and Section.\ref{3}. It was also observed that with an increase in the dimension of subspace, the performance of our algorithm exhibited notable fluctuations. 
	Nonetheless, the error continued to decrease overall. We chose not to further investigate the causes of these fluctuations in this paper, leaving this as an open question for future research.
	\section{Conclusion}
	\par In this paper, we introduced the quantum Krylov-subspace method based linear solver, which was designed to address quantum linear systems problem. Our approach integrated the quantum Krylov-subspace method with the linear combination of unitary operators. It effectively reduce the redundancies of the Fourier approach, subsequently achieved a much better performance in terms of both $\kappa$ and $\epsilon$ in computation complexity. Additionally, we developed a more efficient and general quantum circuit for evaluating the desired overlaps required by the QKSM. On the numerical side, we conducted numerical simulations to evaluate the performance of our proposed algorithm for calculating overlaps on the PennyLane platform. Additionally, we provided a comprehensive analysis of the precision achieved. Furthermore, we simulated both the quantum Krylov-subspace based linear solver and the Fourier approach under ideally condition. With the comparison, we found that our algorithm provided a more precise approximation while using much less quantum computation resources.
	\par In addition to our findings, we also identifies two open questions that merit further exploration and potential investigation in subsequent research:
	\par (1) In our analysis of the algorithm's complexity, we have utilized the convergence of the classical Krylov-subspace method as a provisional upper bound. It remains an open question whether a more precise upper bound can be determined through further analysis.
	\par (2) During the numerical experiment, we noticed that as the quantum Krylov-subspace dimension increases, the performance of our algorithm exhibits noticeable fluctuations. What factors contribute to these variations in performance remains an open question for investigation.
	
	\section*{Acknowledges}
	This work is supported by Key Lab of Guangzhou for Quantum Precision Measurement under Grant No.202201000010, and National Natural Science Foundation of China under Grant No. 12371458.
	\bibliographystyle{IEEEtran}
	\bibliography{Quantum_Krylov-Subspace_Method_Based_Linear_Solver}

\end{document}